\documentclass[aps,prc,twocolumn,superscriptaddress,showpacs,showkeys]{revtex4-1}
\usepackage{graphicx}

\begin{document}
\title{Search for Solar Axions Produced in $p(d,\rm{^3He})A$ Reaction with Borexino Detector}
\author{G.~Bellini,$^1$ J.~Benziger,$^2$ D.~Bick,$^3$ G.~Bonfini,$^4$ D.~Bravo,$^5$ M.~Buizza Avanzini,$^1$ B.~Caccianiga,$^1$
L.~Cadonati,$^6$ F.~Calaprice,$^7$ C.~Carraro,$^8$ P.~Cavalcante,$^4$ A.~Chavarria,$^7$ D.~D$^,$Angelo,$^1$ S.~Davini,$^8$
A.~Derbin,$^9$ A.~Etenko,$^{10}$ K.~Fomenko,$^{11,4}$ D.~Franco,$^{12}$ C.~Galbiati,$^7$ S.~Gazzana,$^4$ C.~Ghiano,$^4$
M.~Giammarchi,$^1$ M.~Goeger-Neff,$^{13}$ A.~Goretti,$^7$ L.~Grandi,$^7$ E.~Guardincerri,$^8$  S.~Hardy,$^5$  Aldo Ianni,$^4$
Andrea Ianni,$^7$ A.~Kayunov,$^9$ D.~Korablev,$^{11}$ G.~Korga,$^4$ Y.~Koshio,$^4$ D.~Kryn,$^{12}$ M.~Laubenstein,$^{4}$
L.~Lewke,$^{13}$ E.~Litvinovich,$^{10}$ B.~Loer,$^7$ F.~Lombardi,$^4$ P.~Lombardi,$^1$ L.~Ludhova,$^1$ I.~Machulin,$^{10}$
S.~Manecki,$^5$ W.~Maneschg,$^{14}$ G.~Manuzio,$^8$  Q.~Meindl,$^{13}$ E.~Meroni,$^1$ L.~Miramonti,$^1$ M.~Misiaszek,$^{15,4}$
D.~Montanari,$^{4,7}$ P.~Mosteiro,$^7$ V.~Muratova,$^9$ L.~Oberauer,$^{13}$ M.~Obolensky,$^{12}$ F.~Ortica,$^{16}$ K.~Otis,$^6$
M. Pallavicini,$^8$ L. Papp,$^5$ L. Perasso,$^1$ S. Perasso,$^8$ A. Pocar,$^6$ R.S. Raghavan,$^5$ G. Ranucci,$^1$ A. Razeto,$^4$
A.~Re,$^1$ P.A.~Romani,$^{16}$ A.~Sabelnikov,$^{10}$ R.~Saldanha,$^7$ C.~Salvo,$^8$ S.~Sch\"{o}nert,$^{13}$ H.~Simgen,$^{14}$
M.~Skorokhvatov,$^{10}$ O.~Smirnov,$^{11}$ A.~Sotnikov,$^{11}$ S.~Sukhotin,$^{10}$ Y.~Suvorov,$^4$ R.~Tartaglia,$^4$
G.~Testera,$^8$ D.~Vignaud,$^{12}$ R.B.~Vogelaar,$^5$ F. von Feilitzsch,$^{13}$ J.~Winter,$^{13}$ M.~Wojcik,$^{15}$
A.~Wright,$^7$ M.~Wurm,$^3$ J.~Xu,$^7$ O.~Zaimidoroga,$^{11}$ S.~Zavatarelli,$^8$ and G.~Zuzel$^{15}$}

\begin{abstract}
\begin{center}
\centering{\textit{\textsuperscript{1}}\textit{ Dipartimento di Fisica, Universita{\textquoteright} degli Studi e INFN, 20133
Milano, Italy}}

\centering{\textit{\textsuperscript{2}}\textit{ Chemical Engineering Department, Princeton University, Princeton, NJ 08544,
USA}}

\centering{\textit{\textsuperscript{3}}\textit{ Institut fur Experimentalphysik, Universitat, 22761 Hamburg, Germany}}

\centering{\textit{\textsuperscript{4}}\textit{ INFN Laboratori Nazionali del Gran Sasso, SS 17 bis Km 18+910, 67010 Assergi
(AQ), Italy}}

\centering{\textit{\textsuperscript{5}}\textit{ Physics Department, Virginia Polytechnic Institute and State University,
Blacksburg, VA 24061, USA}}

\centering{\textit{\textsuperscript{6}}\textit{Physics Department, University of Massachusetts, Amherst, AM01003, USA}}

\centering{\textit{\textsuperscript{7}}\textit{ Physics Department, Princeton University, Princeton, NJ 08544, USA}}

\centering{\textit{\textsuperscript{8}}\textit{ Dipartimento di Fisica, Universita{\textquoteright} e INFN, Genova 16146,
Italy}}

\centering{\textit{\textsuperscript{9}}\textit{St. Petersburg Nuclear Physics Institute, 188350 Gatchina, Russia}}

\centering{\textit{\textsuperscript{10}}\textit{ NRC Kurchatov Institute, 123182 Moscow, Russia}}

\centering{\textit{\textsuperscript{11}}\textit{ Joint Institute for Nuclear Research, 141980 Dubna, Russia}}

\centering{\textit{\textsuperscript{12}}\textit{ Laboratoire AstroParticule et Cosmologie, 75231 Paris cedex 13, France}}

\centering{\textit{\textsuperscript{13}}\textit{ Physik Department, Technische Universitaet Muenchen, 85747 Garching, Germany}}

\centering{\textit{\textsuperscript{14}}\textit{ Max-Plank-Institut fuer Kernphysik, 69029 Heidelberg, Germany}}

\centering{\textit{\textsuperscript{15}}\textit{M. Smoluchowski Institute of Physics, Jagellonian University, 30059 Krakow,
Poland}}

\centering{\textit{\textsuperscript{16}}\textit{Dipartimento di Chimica, Universita{\textquoteright} e INFN, 06123 Perugia,
Italy}}

{Borexino collaboration}
\end{center}

A search for 5.5-MeV solar axions produced in the $p+d\rightarrow\rm{^3He}+A~(5.5\rm{ ~MeV})$ reaction was performed using the Borexino detector. The
Compton conversion of axions to photons, ${\rm A}+e\rightarrow e+\gamma$; the axio-electric effect, ${\rm A}+e+Z\rightarrow e+Z$; the decay of axions
into two photons, ${\rm A}\rightarrow2\gamma$; and inverse Primakoff conversion on nuclei, ${\rm A}+Z\rightarrow\gamma+Z$, are considered. Model
independent limits on axion-electron ($g_{Ae}$), axion-photon ($g_{A\gamma}$), and isovector axion-nucleon ($g_{3AN}$) couplings are obtained:
$|g_{Ae}\times g_{3AN}| \leq 5.5\times 10^{-13}$ and $|g_{A\gamma}\times g_{3AN}| \leq 4.6\times 10^{-11} \rm{GeV}^{-1}$ at $m_A <$ 1 MeV (90\%
c.l.). These limits are 2-4 orders of magnitude stronger than those obtained in previous laboratory-based experiments using nuclear reactors and
accelerators.
\end{abstract}


\maketitle

\section{INTRODUCTION}

The axion hypothesis was introduced by Weinberg \cite{Wei78} and Wilczek \cite{Wil78}, who showed that the solution to the
problem of $CP$ conservation in strong interactions, proposed earlier by Peccei and Quinn \cite{Pec77}, should lead to the
existence of a neutral pseudoscalar particle. The original WWPQ axion model produced specific predictions for the coupling
constants between axions and photons ($g_{A\gamma}$), electrons ($g_{Ae}$), and nucleons ($g_{AN}$) which were soon disproved by
experiments performed with reactors and accelerators, and by experiments with artificial radioactive sources \cite{PDG10}.

Two classes of new theoretical models, hadronic or KSVZ \cite{Kim79,Shi80} and GUT or DFSZ   \cite{Zhi80,Din81}, describe
"invisible" axions, which solve the $CP$ problem in strong interactions and interact more weakly with matter. The scale of
Peccei-Quinn symmetry violation ($f_A$) in both models is arbitrary and can be extended to the  Planck mass $m_P \approx
10^{19}$ GeV. The axion mass in these models is determined by the axion decay constant $f_A$:
\begin{equation}\label{ma}
  m_A\approx (f_\pi m_\pi /f_A) (\sqrt{z}/(1+z)),
\end{equation}
where $m_\pi$ and $f_\pi$ are, respectively, the mass and decay constant of the neutral $\pi$ meson and  $z = m_u/m_d$ is $u$ and $d$ quark-mass
ratio.  The equation (\ref{ma}) can be rewritten as: $m_A(\rm{eV})\approx 6.0\times 10^6/\it{f_A} {\rm{(GeV)}}$. Since the axion–-hadron and
axion–-lepton interaction amplitudes are proportional to the axion mass, the interaction between axions and matter is suppressed.

The effective coupling constants $g_{A\gamma}$, $g_{Ae}$, and $g_{AN}$  are to a great extent  model dependent. For example, the hadronic axion
cannot interact directly with leptons, and the constant $g_{Ae}$ exists only because of radiative corrections. Also, the constant $g_{A\gamma}$ can
differ by more than two orders of magnitude from the values accepted in the KSVZ and DFSZ models \cite{Kap85}.

The results from present-day experiments are interpreted within these two most popular axion models. The main experimental efforts are focused on
searching for an axion with a mass in the range of $10^{-6}$ to $10^{-2}$ eV. This range is free of astrophysical and cosmological constraints, and
relic axions with such a mass are considered to be the most likely dark matter candidates.

New solutions to the $CP$ problem rely on the hypothesis of a world of mirror particles \cite{Bere00,Bere01} and super-symmetry
\cite{Hal04}. These models allow the existence of axions with a mass of about 1 MeV, which are not precluded by laboratory
experiments or astrophysical data.

The purpose of this study is to search experimentally for solar axions with an energy of 5.5 MeV,  produced in the $p + d \rightarrow\rm{^3He}+ A$
(5.49 MeV) reaction. The axion flux is thus proportional to the $pp$-neutrino flux, which is known with a high accuracy \cite{Ser09, Bel11A}. The
range of axion masses under study has been extended to 5 MeV. The axion detection signatures exploited in this study are Compton axion to photon
conversion, ${\rm A}+e\rightarrow e+\gamma$, and the axio-electric effect, ${\rm A}+e+Z\rightarrow e+Z$. The amplitudes of these processes are
defined by the $g_{Ae}$ coupling. We also consider the potential signals from axion decay into two $\gamma$-quanta and from inverse Primakoff
conversion on nuclei, ${\rm A}+Z\rightarrow\gamma+Z$. The amplitudes of these reactions depend on the axion-photon coupling $g_{A\gamma}$. The
signature of all these reactions is a 5.5 MeV peak.


We have previously published a search for solar axions emitted in the 478 keV M1-transition of $^7{\rm{Li}}$  using the Borexino counting test facility
\cite{Bel08}.

The results of laboratory searches for the axion as well as astrophysical and cosmological axion bounds can be found in
\cite{PDG10}.

\section{ The flux of 5.5 MeV axions}
The Sun potentially represents an efficient and intense source of axions. One production mechanism is photon–-axion conversion
in the electromagnetic fields of the solar plasma. In addition, electrons could produce axions via Compton processes and
bremsstrahlung.  Finally, monochromatic axions could be emitted in magnetic transitions in nuclei, when low-lying levels are
thermally excited by the high temperature of the Sun.

Even the reactions of the $\emph{pp}$-solar fusion chain and the CNO cycle can produce axions. The most intense flux is expected
from the formation of the $\rm{^3He}$ nucleus:
\begin{equation}
 p + d\rightarrow \rm{{^3He} + \gamma ~(5.5\;MeV)}.
\end{equation}

According to the Standard Solar Model (SSM), 99.7\% of all deuterium is produced from the fusion of two protons, $p + p \rightarrow d + e^+ + \nu_e$,
while the remaining 0.3\% is due to the  $p+ p + e^- \rightarrow  d + \nu_e$ reaction. The produced deuteron captures a proton with lifetime $\tau =
6 s$.

The expected solar axion flux can thus be expressed in terms of the $pp$-neutrino flux. The proportionality factor between the axion and neutrino
fluxes is determined by a dimensionless axion-nucleon coupling constant $g_{AN}$, which consists of isoscalar $g_{0AN}$ and isovector $g_{3AN}$
components. The ratio between  the probability of an M1 magnetic nuclear transition with axion production $(\omega_A)$ and photon production
$(\omega_\gamma)$ can be expressed as \cite{Hax91}-\cite{Avi88}:

\begin{equation}\label{axion_prob}
\frac{\omega_{A}}{\omega_{\gamma}} =
\frac{1}{2\pi\alpha}\frac{1}{1+\delta^2}\left[\frac{g_{0AN}\beta_{1}+g_{3AN}}{(\mu_{0}-0.5)\beta_1+\mu_{3}-\eta_1}\right]^{2}
\left(\frac{p_{A}}{p_{\gamma}}\right)^{3},
\end{equation}
where $p_{\gamma}$ and $p_{A}$ are, respectively, the photon and axion momenta; $\delta^2 = E/M$ is the ratio between the
probabilities of $E$ and $M$ transitions; $\alpha\approx 1/137$ is the fine-structure constant; $\mu_0 = \mu_p + \mu_n\approx
0.88$ and $\mu_3 = \mu_p - \mu_n \approx 4.71$ are, respectively, the isoscalar and isovector nuclear magnetic moments; and
$\beta_1$ and $\eta_1$ are parameters dependent on the specific nuclear matrix elements.

Within the hadronic axion model, the constants $g_{0AN}$ and $g_{3AN}$ can be written in terms of the axion mass
\cite{Kap85},\cite{Sre85}:
\begin{eqnarray}\label{gan0}
\nonumber g_{0AN}=-\frac{m_N}{6f_A}[2S_{fs}+(3F-D)\frac{1+z-2w}{1+z+w}]=\\
= -4.03\times 10^{-8}(m_A/1 {\rm{eV}}),\label{gan0_2}
\end{eqnarray}
\begin{eqnarray}\label{gan3}
\nonumber g_{3AN}= -\frac{m_N}{2f_A}[(D+F)\frac{1-z}{1+z+w}]=\\
=-2.75 \times 10^{-8}(m_A/1 {\rm{eV}}).\label{gan3_2}
\end{eqnarray}
where $m_N\approx939$ MeV is the nucleon mass, and $z = m_u/m_d \cong 0.56$ and $w = m_u/m_s \cong0.029$ are $u$, $d$ and $s$
quark-mass ratios. Axial-coupling parameters $F$ and $D$ are obtained from hyperon semi-leptonic decays with high precision:
$F$=0.462 $\pm$ 0.011, $D$= 0.808 $\pm$ 0.006 \cite{Mat05}. The parameter $S_{fs}$, characterizing the flavor singlet coupling
is poorly constrained: $(0.37\leq S_{fs}\leq0.53)$ and $(0.15\leq S_{fs}\leq0.5)$ were found in \cite{Alt97} and \cite{Ada97},
respectively. The values of the axion-nucleon couplings given in (\ref{gan0_2}) and (\ref{gan3_2}) are obtained assuming
$S_{fs}$=0.5. The value of $u$- and $d$-quark-mass ratio $z$ = 0.56 is generally accepted for axion papers, but it could vary in
the range ($0.35-0.6$) \cite{PDG10}. These uncertainties in $S_{fs}$ and $z$ could cause the values of $g_{0AN}$ and $g_{3AN}$
to differ from (\ref{gan0_2}) and (\ref{gan3_2}) by factors of (0.4--1.3) and (0.9--1.9) times, respectively.

The values of $g_{0AN}$ and $g_{3AN}$ in the DFSZ model depend on an additional unknown parameter, but have the same order of
magnitude: they have ($0.3 - 1.5$) times the values of the corresponding constants for the hadronic axion.
\begin{figure}
\includegraphics[width=9cm,height=10.5cm]{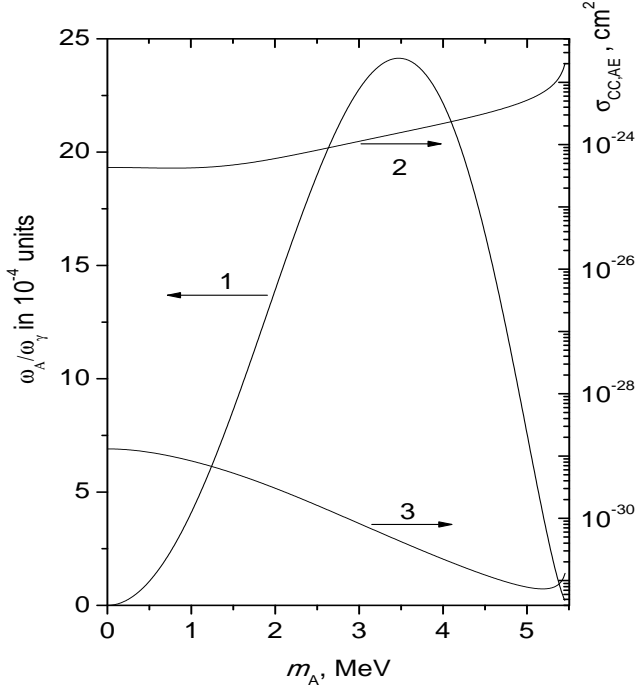}
\caption {Ratio of the emission probabilities for axions and $\gamma$ quanta ($\omega_A/\omega_\gamma$) in the $p + d \rightarrow {^3{\rm{He}}} + \gamma$
reaction (curve 1, left-hand scale); cross section of the Compton conversion and axio-electric effect for 5.5-MeV axions on carbon atoms for $g_{Ae} =
1$ (curve 2 and 3,  right-hand scale). } \label{Fugure:Ratio_CS}
\end{figure}

In the $p+d\rightarrow \rm{^3He}+\gamma$ reaction, the M1-type transition corresponds to the capture of a proton with zero
orbital momentum. The probability, $\chi$, of proton capture from the $S$ state at energies below 80 keV was measured in
\cite{Sch97}; at a proton energy of $\sim 1$ keV, $\chi$ = 0.55 $(\delta^2 = 0.82)$. The proton capture from the $S$ state
corresponds is an isovector transition, and the ratio $\omega_A/\omega_\gamma$, from expression (3), therefore depends only on
$g_{3AN}$ \cite{Don78}:
\begin{equation}\label{ratio}
\frac{\omega_{A}}{\omega_{\gamma}} =
 \frac{\chi}{2\pi\alpha}\left[\frac{g_{3AN}}{\mu_3}\right]^2\left(\frac{p_A}{p_\gamma}\right)^3 = 0.54(g_{3AN})^2
 \left(\frac{p_A}{p_\gamma}\right)^3.
\end{equation}

The calculated values of the $\omega_A/\omega_\gamma$ ratio as a function of the axion mass are shown in Fig.\ref{Fugure:Ratio_CS}. The expected
solar axion flux on the Earth's surface is then
\begin{eqnarray}\label{FluxA}
\Phi_{A0} = \Phi_{\nu pp}(\omega_A/\omega_\gamma) = 3.23\times10^{10}(g_{3AN})^2(p_A/p_\gamma)^3,
\end{eqnarray}
where $\Phi_{\nu p p} = 6.0 \times 10^{10} {\rm{cm}}^{-2} {\rm{s}}^{-1}$ is the $pp$ solar neutrino flux \cite{Ser09, Bel11A}.
Using the relation between $g_{3AN}$ and $m_A$ given by (\ref{gan3}), the $\Phi_{A0}$ value appears to be proportional to
$m_A^2$: $\Phi_{A0} = 2.44\times10^{-5}m_A^2(p_A/p_\gamma)^3$, where $m_A$ is given in eV units.


\section{ INTERACTION OF AXIONS WITH MATTER AND AXION DECAYS}

\subsection{Axion-electron interactions: Compton conversion and the axio-electric effect}

An axion can scatter an electron to produce a photon in the Compton-like process $A+e\rightarrow\gamma+e$. The Compton
differential cross section for electrons was calculated in  \cite{Don78}, \cite{Avi88}, \cite{Zhi79}. The energy spectrum of the
$\gamma$-quanta depends on the axion mass, while the spectra of electrons can be found from relation $E_e = E_A-E_{\gamma}$.
Here, $E_A\cong$ 5.49 MeV, which is the Q-value of the $p(d,\rm{^3He})\gamma$ reaction. The integral cross section corresponding
to this mode is  \cite{Don78}, \cite{Avi88}, \cite{Zhi79}:
\begin{eqnarray}
\nonumber \sigma_{CC}=\frac{g_{Ae}^{2}\alpha}{8m^{2}p_{A}}[\frac{2m^{2}(m+E_{A})y}{(m^{2}+y)^{2}} +\\
\nonumber +\frac{4m(m_{A}^{4}+2m_{A}^{2}m^{2}-4m^{2}E_{A}^{2})}{y(m^{2}+y)} +\\
+\frac{4m^{2}p_{A}^{2}+m_{A}^{4}}{p_{A}y}\ln\frac{m+E_{A}+p_{A}}{m+E_{A}-p_{A}}].\label{CC_CS}
\end{eqnarray}
where $p_{A}$ and $E_{A}$ are the momenta and the energy of the axion respectively and $y=2mE_{A}+m_{A}^{2}$. The dimensionless coupling constant
$g_{Ae}$ is associated with the electron mass $m$, so that $g_{Ae}=C_em/f_{A}$, where $C_e$ is a model dependent factor of the order of unity. In the
standard WWPQ axion model, the values $f_A$=250 GeV and $C_e$=1 are fixed and $g_{Ae}\approx 2\times10^{-6}$. In the DFSZ axion models
$C_e=1/3\cos^2\beta_{\rm{dfsz}}$, where $\beta_{\rm{dfsz}}$ is an arbitrary angle. Assuming $\cos^2\beta_{\rm{dfsz}}$=1, the axion-electron coupling
is $g_{Ae}$=2.8$\times10^{-11}m_A$ where $m_A$ is expressed in $eV$ units. The hadronic axion has no tree-level couplings to the electron, but there
is an induced axion-electron coupling at one-loop level \cite{Sre85}:
\begin{equation}
g_{Ae}=\frac{3n\alpha^{2}m}{2\pi f_{a}}\left(\frac{E}{N}\ln\frac{f_{A}}{m}-\frac{2}{3}\frac{4+z+w}{1+z+w}\ln\frac{\Lambda}{m}\right)\label{Gaee}
\end{equation}
where $n$ is the number of generations, $N$ and $E$ are the model dependent coefficients of the color and electromagnetic anomalies and
$\Lambda\approx$1 GeV is the cutoff at the QCD confinement scale. The interaction strength of the hadronic axion with the electron is suppressed by a
factor $\sim\alpha^{2}$.

The integral cross section $\sigma_{CC}$ calculated for $g_{Ae} = 1$  is shown in Fig.\ref{Fugure:Ratio_CS}. For axions with fixed $g_{Ae}$ (curve 2
in Fig. \ref{Fugure:Ratio_CS}), the phase space contribution to the cross section is approximately independent of $m_{A}$ for $m_{A}<$ 2 MeV and the
integral cross section is:
\begin{equation}
\sigma_{CC}\approx g_{Ae}^{2}\times4.3\times10^{-25} \rm{cm^{2}}.\label{CC_CS_Aprox}
\end{equation}

The other process associated with axion-electron coupling is the axio-electric effect $A+e+Z\rightarrow e+Z$ (the analogue of the photo-electric
effect). In this process the axion disappears and an electron is emitted from an atom with an energy equal to the energy of the absorbed axion minus
the electron binding energy $E_{b}$. The cross section of the axio-electric effect on K-electrons where the axion energy $E_{A}\gg E_{b}$ was
calculated in \cite{Zhi79} and has a complex form; it is shown in Fig. \ref{Fugure:Ratio_CS}. The cross section has a $Z^{5}$ dependence and for
carbon atoms the cross section is $\sigma_{Ae}\approx g_{Ae}^{2}\times$1.3$\times$10$^{-29}$ cm$^{2}$/electron for $m_A <$ 1 MeV. This value is more
than 4 orders of magnitude lower than for axion Compton conversion. However, thanks to the different energy dependence ($\sigma_{CC}\sim E_{A}$,
$\sigma_{Ae}\sim(E_{A})^{-3/2}$) and $Z^{5}$ dependence, the axio-electric effect is a potential signature for axions with detectors having high $Z$
active mass \cite{Der10}.

For axions with a mass above $2m$, the main decay mode is the decay into an electron-positron pair: $A \rightarrow e^+ + e^-$. The lifetime of an
axion in the intrinsic reference system has the form:
\begin{equation}
\tau_{e^+e^-}=8\pi/(g^2_{Ae}\sqrt{m^2_A - 4m^2_e}).
\end{equation}
The probability of an axion to reach the Earth is
\begin{equation}
P(m_A,p_A)=\exp(-\tau_{f}/\tau_{e^+e^-}),
\end{equation}
where $\tau_{f}$ is the time of flight in the reference system associated with the axion:
\begin{equation}
\tau_{f}=\frac{Lm_A}{cp_A}= \frac{m_{A}}{E_{A}}\frac{L}{\beta c}\label{tauf}.
\end{equation}
Here $L = 1.5\times 10^{13}$ cm is the distance from the Earth to the Sun and $\beta=p_{A}/E_{A}$ is the axion velocity in terms
of the speed of light. The condition $\tau_{f}<0.1\tau_{e^+e^-}$ (in this case, $90\%$ of all axions reach the Earth) limits the
sensitivity of solar axion experiments to $g_{Ae} < (10^{-12}-10^{-11})$ \cite{Der10}.

\subsection{Axion-photon interaction: axion decay and the inverse Primakoff conversion on nuclei}
If the axion mass is less than $2m$, $A\rightarrow e^+ + e^-$ decay is forbidden, but the axion can decay into two $\gamma$ quanta. The probability
of the decay, which depends on the axion–-photon coupling constant and the axion mass, is given by the expression:
\begin{equation}
\tau_{2\gamma}=\frac{64\pi}{g_{A\gamma}^2m_A^3}.\label{tau_CM}
\end{equation}
where $g_{A\gamma}$ is an axion-photon coupling constant with dimension of (energy)$^{-1}$ which is presented as in
\cite{Kap85},\cite{Sre85}:
\begin{equation}
g_{A\gamma}=\frac{\alpha}{2\pi f_{A}}\left(\frac{E}{N}-\frac{2(4+z+w)}{3(1+z+w)}\right)\equiv\frac{\alpha}{2\pi
f_{A}}C_{A\gamma\gamma}\label{C_gamma}
\end{equation}
where $E/N$ is a model dependent parameter of the order of unity. $E/N$ = 8/3 in the DFSZ axion models
($C_{A\gamma\gamma}$=0.74) and $E/N$ = 0 for the original KSVZ axion ($C_{A\gamma\gamma}$=-1.92).

The phase space for decay depends on $m_A^3$. For $\tau_{2\gamma}$ measured in seconds, $g_{A\gamma}$ in ${\rm GeV^{-1}}$, and $m_A$ in ${\rm eV}$,
one obtains:
\begin{equation}
\tau_{2\gamma}=1.3\times 10^5 g_{A\gamma}^{-2}m_A^{-3} = 3.5\times 10^{24}m_A^{-5}C_{A\gamma\gamma}^{-2}. \label{axion_decay}
\end{equation}
The flux of axions reaching the detector is given by
\begin{equation}
\Phi_{A}={\rm{exp}}(-\tau_{f}/\tau_{2\gamma})\Phi_{A0} = {\rm{exp}}(-\tau_{f} g_{A\gamma}^2 m_A^3/64\pi)\Phi_{A0} \label{Flux}
\end{equation}
where $\Phi_{A0}$ is the axion flux at the at the Earth in case there is no axion decay (\ref{FluxA}), $\tau_{2\gamma}$ is defined by (\ref{tau_CM},
\ref{axion_decay}), and $\tau_{f}$, given by (\ref{tauf}) is the time of flight in the axion frame of reference. Because of axion decay, the
sensitivity of experiments using solar axions drops off for large values of $g_{A\gamma}^{2}m_A^{3}$.

\begin{figure}
\includegraphics[width=9cm,height=10.5cm]{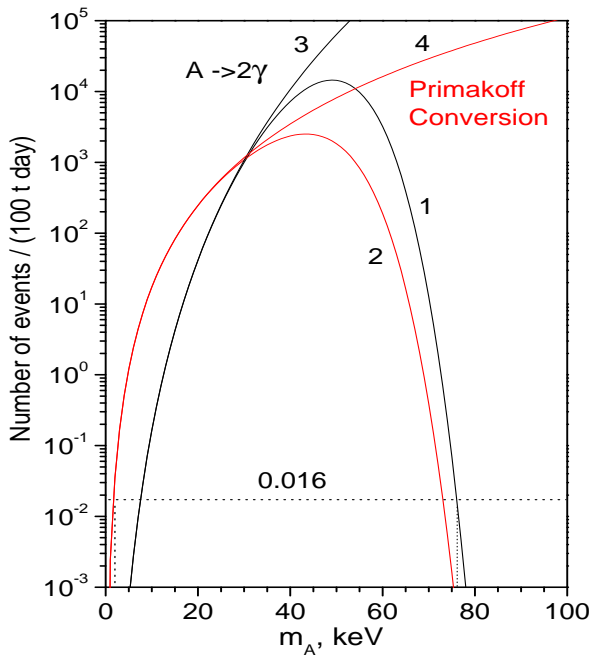}
\caption {The expected number of axion decays (l) and inverse Primakoff conversions on $^{12}\rm{C}$ nuclei (2) in the 100 t  per day for KSVZ axion
model. Lines (3) and (4) show the corresponding curves under the assumption that axions do not decay during their flight from the Sun.}
\label{Figure:Decay_PC}
\end{figure}

The number of $A\rightarrow 2\gamma$ decays in a detector of volume V is:
\begin{equation}
N_{\gamma}=\Phi_{A}\frac{Vm_{A}}{\beta cE_{A}\tau_{2\gamma}}\label{N_gamma}.
\end{equation}
This leads, using the KSVZ model, to expected Borexino event rates like those shown in Fig.\ref{Figure:Decay_PC} for different
values of $m_{A}$.  As can be seen in the Figure, the expected event rate is peaked, with a drop-off at low $m_{A}$ due to the
lower axion decay rate in the detector, and a decrease at high $m_{A}$ resulting from the reduced flux from axion decay in
flight.   The maximum $N_{\gamma}$ corresponds to $m_{A}$ = $((8/6)\tau_{2\gamma}m_{A}^{5}/(\tau_{f}/m_{A}))^{1/6}$= 65 keV,
where $\tau_{2\gamma}$ and $\tau_{f}$ are defined by (\ref{axion_decay}) and (\ref{tauf}).

Another process depending on $g_{A\gamma}$ coupling is the Primakoff photo-production on carbon nuclei
$A+{^{12}\rm{C}}\rightarrow\gamma+{^{12}\rm{C}}$. The integral inverse Primakoff conversion cross section is \cite{Avi88}:
\begin{equation}
\sigma_{PC}=g_{A\gamma}^{2}\frac{Z^{2}\alpha}{2}\left[\frac{1+\beta^{2}}{2\beta^{2}}\ln\left(\frac{1+\beta}{1-\beta}\right)-\frac{1}{\beta}\right].\label{Primakoff}
\end{equation}
Because the cross section depends on the $g_{A\gamma}$ coupling, the decrease in the axion flux due to $A\rightarrow 2\gamma$
decays during their flight from the Sun should be taken into account. The axion flux at the detector was calculated by the
method described above. The atomic-screening corrections for $^{12}$C were introduced following the method proposed in
\cite{Avi88}. The expected conversion rate in Borexino is shown in Fig.\ref{Figure:Decay_PC} for different values of $m_{A}$.

\subsection{Escape of axions from the Sun}

Axions could be captured within the Sun. The requirement that most axions escape the Sun thus limits the axion coupling
strengths accessible to terrestrial experiments. Each of the 4 axion-matter interactions considered in this paper contribute to
these limits.

The flux of 5.5 MeV axions on the Earth's surface is proportional to the $pp$-neutrino flux, as given in equation (\ref{FluxA}), only when the axion
lifetime exceeds the time of flight from the Sun and when the flux is not reduced as a result of axion absorption by solar matter. Axions produced at
the center of the Sun cross a layer of approximately $6.8\times 10^{35}$ electrons/${\rm{cm}}^{2}$ in order to reach the Sun's surface. Axion loss
due to Compton conversion into photons in the solar matter imposes an upper limit on $g_{Ae}$ after which the sensitivity of terrestrial experiments
using solar axions is reduced. The cross section of the Compton conversion reaction for 5.5-MeV axions depends weakly on the axion mass and can be
written as $\sigma_{CC}\approx g^2_{Ae}\times4.3\times 10^{-25}{\rm{cm}}^2$. For $g_{Ae}$ values below $10^{-6}$, the axion flux is not substantially
suppressed.

The maximum cross section of the axio-electric effect on atoms is $\sigma_{Ae}\approx g^2_{Ae} Z^2 1.9\times
10^{-29}{\rm{cm}}^2$ (see Fig.\ref{Fugure:Ratio_CS} for carbon). The abundance of heavy ($Z > 50$) elements in the Sun is
$\sim10^{-9}$ in relation to hydrogen \cite{Asp06}. If  $g_{Ae}< 10^{-3}$, the change in the axion flux does not exceed 10\%.

The axion-photon interaction, as determined by the constant $g_{A\gamma}$, leads to the conversion of an axion into a photon in
a field of nucleus. The cross section of the reaction is $\sigma_{pc}\approx g^2_{A\gamma}Z^2\times1.8\times
10^{-29}{\rm{cm}}^2$. Taking into account the density of $^1\rm{H}$ and $^4\rm{He}$ nuclei, the condition that axions
efficiently escape the Sun imposes the constraint $g_{A\gamma} < 10^{-4} {\rm{GeV}}^{-1}$. Constraint for the other elements are
negligible due to their low concentration in the Sun.

The axion-nucleon interaction leads to axion absorption in a threshold reaction similar to photo--dissociation: $A+ Z \rightarrow Z_1 + Z_2$.
 For axions with energy 5.5-MeV this can occur for only a few nuclei: $^{17}{\rm{O}}$,$ ^{13}{\rm{C}}$, and $^2{\rm{H}}$. It was shown in \cite{Raf82} that
axio--dissociation cannot substantially reduce the axion flux for $g_{AN} < 10^{-3}$.

In all, the requirement that most axions escape the Sun sets these limits on the matter-axion couplings - $g_{Ae}< 10^{-6}$,
$g_{A\gamma} < 10^{-4} {\rm{GeV}}^{-1}$ and $g_{AN} < 10^{-3}$.

\section{Experimental set-up and measurements}
\subsection {Brief description of Borexino}
Borexino is a real-time detector for solar neutrino spectroscopy located at the Gran Sasso Underground Laboratory. Its main goal
is to measure low energy solar neutrinos via ($\nu$,e)-scattering in an ultra-pure liquid scintillator. At the same time,
however, the extremely high radiopurity of the detector and its large mass allow it to be used to study other fundamental
questions in particle physics and astrophysics.

The main features of the Borexino detector and its components have been thoroughly described in \cite{Ali02}-\cite{Bel12B}. Borexino is a
scintillator detector with an active mass of 278 tons of pseudocumene (C$_9$H$_{12}$), doped with 1.5 g/liter of PPO (C$_{15}$H$_{11}$NO). The
scintillator is housed in a thin nylon vessel (inner vessel - IV) and is surrounded by two concentric pseudocumene  buffers (323 and 567 tons) doped
with a small amount of light quencher (dimethyl phthalate~- DMP) to reduce their scintillation. The two buffers are separated by a second thin nylon
membrane to prevent diffusion of radon coming from PMTs, light concentrators and SSS walls towards the scintillator. The scintillator and buffers are
contained in a Stainless Steel Sphere (SSS) with diameter 13.7 m. The SSS is enclosed in an 18.0-m diameter, 16.9-m high domed Water Tank (WT),
containing 2100 tons of ultra pure water as an additional shield against external $\gamma$'s and neutrons. The scintillation light is detected by
2212 8" PMTs uniformly distributed on the inner surface of the SSS. The WT is equipped with 208 additional PMTs that act as a Cerenkov muon detector
(outer detector) to identify the residual muons crossing the detector. All the internal components of the detector were selected following stringent
radiopurity criteria.

\subsection{Detector calibration. Energy and spatial resolutions.}
In Borexino, charged particles are detected by scintillation light induced by their interactions with the liquid scintillator.
The energy of an event is related to the total collected light by the PMTs. In a simple approach, the response of the detector
is assumed to be linear with respect to the energy released in the scintillator. The coefficient linking the event energy and
the total collected charge is called the light yield (or photo-electron yield). Deviations from linearity at low energies can be
taken into account including the ionization deficit function $ f(k_{B},E) $, where $k_B$ is the empirical Birks' constant.

The detector energy and spatial resolution were studied with radioactive sources placed at different positions inside the inner
vessel. For relatively high  energies ( $>$2 MeV), which are of interest for 5.5 MeV axion studies, the energy calibration was
performed with a $^{241}$Am-$^9$Be neutron source. One can find a detailed description of the energy calibration in
\cite{Bel10,Bel10A}. Deviations of the $\gamma$-peak positions from linearity was less than 30 keV over the whole energy range.
The energy resolution scales approximately as ${(\sigma /E)}$ $\simeq (0.058+1.1\times10^{-3}E)/\sqrt{E}$ where E is given in
MeV units. The position of an event is determined using a photon time of flight reconstruction algorithm. The resolution of the
event reconstruction, as measured using the $^{214}$Bi-$^{214}$Po $\beta-\alpha$ decay sequence, is 13$\pm$2 cm \cite{Ali09}.

\subsection{Data selection}

\begin{figure}
\includegraphics[bb = 30 90 500 760, width=8cm,height=10cm]{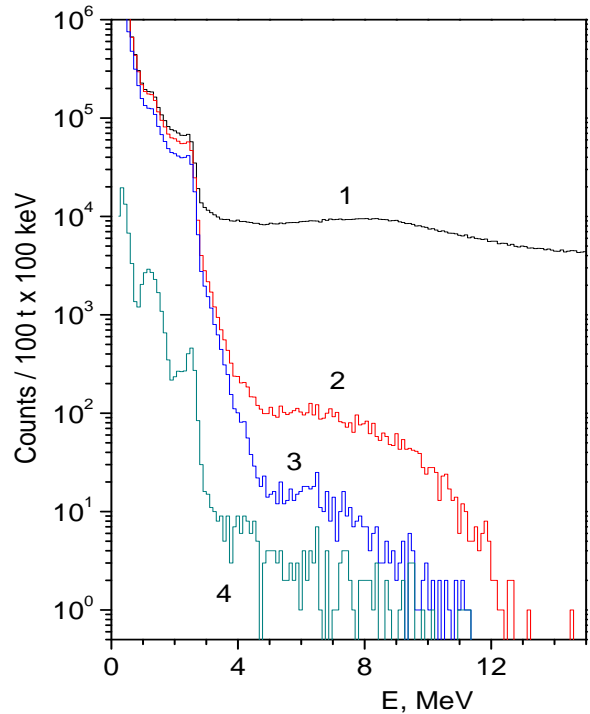}
\caption { Energy spectra of the events and effect of the selection cuts. From top to bottom: (1) raw spectrum; (2) with 2 ms muon veto cut; (3) with
events within 6.5 s of a muon crossing the SSS removed; (4) events inside FV } \label{Figure:Spectra}
\end{figure}

The experimental energy spectrum from Borexino in the range (1.0-15) MeV, containing 737.8~live-days of data, is shown in
Fig.\ref{Figure:Spectra}.  At energies below 3~MeV, the spectrum is dominated by 2.6 MeV $\gamma$'s from the $\beta$-decay of
$^{208}$Tl in the PMTs and in the SSS.

The spectrum obtained by vetoing all muons and events within 2 ms after each muon is shown by curve 2, Fig.\ref{Figure:Spectra}. Muons are rejected
by the outer detector and by an additional cut on the mean time of the hits belonging to the cluster and on the time corresponding to the maximum
density of hits. This cut rejects residual muons that were not tagged by the outer water Cherenkov detector and that interacted in the pseudocumene
buffer regions (see \cite{Bel11B} for more details).

To reduce the background due to short-lived isotopes (1.1 s $^8\rm{B}$, 1.2 s $^8\rm{Li}$, etc; see \cite{Bel10A}) induced by
muons, an additional 6.5 s veto is applied after each muon crossing the SSS (curve 3, Fig.\ref{Figure:Spectra}). This cut
induces 202.2 days of dead time that reduces the live-time to 535.6 days.

In order to reject external background in the 5.5 MeV energy region a fiducial volume cut is applied. Curve 4 of
Fig.\ref{Figure:Spectra} shows the effect of selecting a 100 ton fiducial volume (FV) by applying a cut  R $\leq$ 3.02 m.
Additionally, a pulse shape-discrimination analysis based on the Gatti optimal filter \cite{Gat62} is performed: events with
negative Gatti variable corresponding to $\gamma$- and $\beta$-like signals are selected (see \cite{Ali09} for more details).
This cut does not change the spectrum for energies higher than 4 MeV.

\subsection{Simulation of the Borexino response functions}
\begin{figure}
\begin{center}\includegraphics[width=10cm, height=10cm]{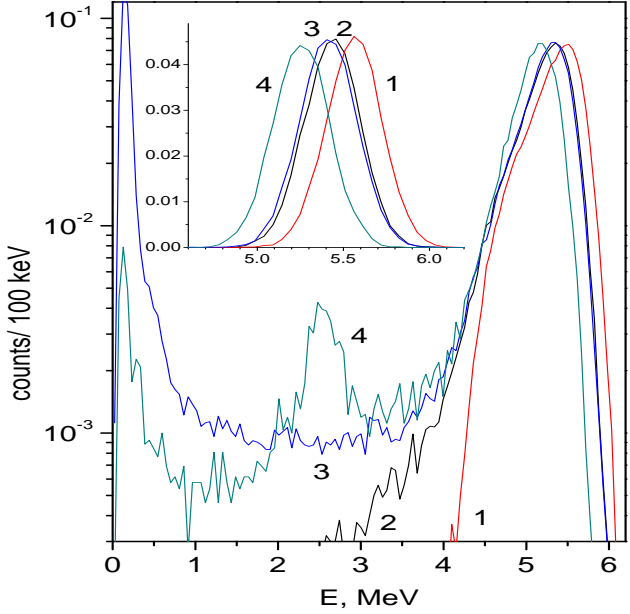}\end{center}
\caption{\label{Figure:BorexinoResponse}Simulated responses to axion interactions in the Borexino IV: 1- axio-electric effect (
5.49 MeV electrons), 2- Compton axion to photon conversion (electrons and $\gamma$-quanta), 3- Primakoff conversion (5.49 MeV
$\gamma$-quanta), 4- decay $A\rightarrow 2\gamma$. The inset shows the corresponding responses for events reconstructed within
the FV.} \label{Figure:Response_Functions}
\end{figure}

The Monte Carlo (MC) method has been used to simulate the Borexino response $S(E)$ to electrons and $\gamma$-quanta produced by
axion interactions. The MC simulations are based on the GEANT4 code, taking into account the effect of ionization quenching and
non-linearity induced by the energy dependence on the event position. Uniformly distributed $\gamma$'s were simulated inside the
entire inner vessel, but only those which reconstructed within the FV were used in determining the response function. The MC
candidate events were selected by the same cuts applied in the real data selection.

The energy spectra of electrons and gammas from the axion Compton conversion were generated according to the differential cross
section given in \cite{Don78}, \cite{Avi88}, \cite{Zhi79} for different axion masses \cite{Bel08}. The responses for the axion
decay into two $\gamma$ quanta were calculated taking into account the angular correlation between photons. The response
functions for axion Compton conversion (electron and $\gamma$-quanta with total energy of 5.5 MeV), for the axio-electric effect
(electron with energy 5.5 MeV), axion decay (two $\gamma$-quanta with energy 2.75 MeV in case of non-relativistic axions) and
for Primakoff conversion (5.5 MeV $\gamma$-quanta) are shown in Fig.\ref{Figure:Response_Functions}. The response functions are
normalized to 1 axion interaction (decay) in the IV. The shift in the position of the total absorption peak for interactions
involving $\gamma$'s is caused by an ionization quenching effect. All response functions are fitted with Gaussians.

\section{Results and discussions}
\subsection{Fitting procedure}

\begin{figure}
\includegraphics[width=9cm,height=10.5cm]{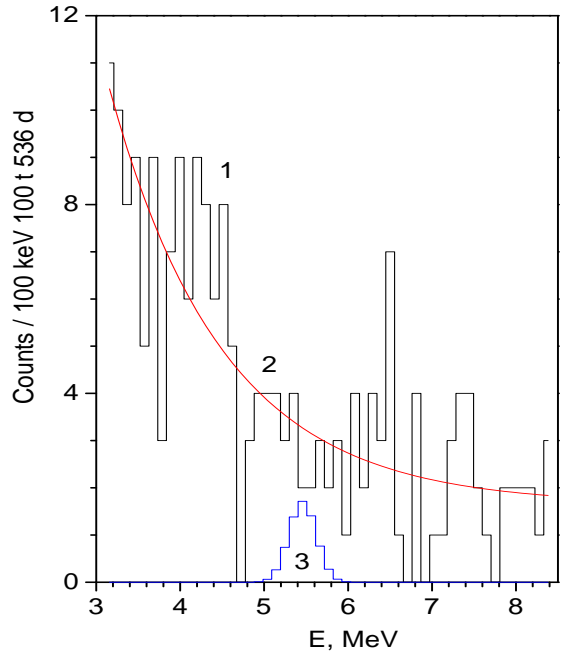}
\caption {The fitted Borexino spectrum in the ($3.2-8.4$) MeV range. Curve 3 is the detector response function for Compton
axion-photon conversion at the 90\% c.l. upper limit ($S$=6.9 events). } \label{Figure:Fit_Compton}
\end{figure}

Figure \ref{Figure:Fit_Compton}  shows the observed Borexino energy
spectrum in the ($3.0-8.5$) MeV range in which the axion peaks might appear. The spectrum is modeled with a sum
of exponential and Gaussian functions,
\begin{equation}
N^{\rm th}(E) = a+b\times e^{-cE}+(S/\sqrt{2\pi}\sigma)\times e^{-\frac{(E_{MC}-E)^2}{2\sigma^2}},\label{Function:fit}
\end{equation}
where the position $E_{MC}$ ($\cong$5.49 MeV) and dispersion $\sigma$ ($\cong$0.15 MeV) are taken from the MC response, $S$ is
the peak intensity and $a, b$ and $c$ are the parameters of the function describing the continuous background.

The number of events in the axion peak $S$ was calculated using the maximum likelihood method. The likelihood function assumes the form of a product
of Poisson probabilities:
\begin{equation}
 L=\prod e^{-N^{\rm th}_i}({N^{\rm th}_i})^{N^{\rm exp}_{i}} / N^{\rm exp}_{i}!\label{Function:Poisson}
\end{equation}
where $N^{\rm th}_i$ and $N^{\rm exp}_i$ are the expected (\ref{Function:fit}) and measured number of counts in the i-th bin of the spectrum,
respectively. The dispersion of the peak ($\sigma$) was fixed, while the position ($E_0$) was varied around $E_{MC}\pm30$ keV, to take into account
the uncertainty in the energy scale. The others 4 parameters ($a, b, c$ and $S$) were also free. The total number of the degrees of freedom in the
range of 3.2–-8.4 MeV was 46.

The fit results, corresponding to the maximum of $L$  at $S$=0 are shown in Fig.\ref{Figure:Fit_Compton}. The value of modified $\chi^2 = \sum
(N^{\rm exp}_i-N^{\rm th}_i)^2/N^{\rm th}_i$ is $\chi^2$= 44/46. Because of the low statistics, a  Monte Carlo simulation of (\ref{Function:fit}) is
used to find the probability of $\chi^2_p \geq$ 44. The goodness-of-fit ($ p = 52\%$) shows that the background is well described by function
(\ref{Function:fit}). The upper limit on the number of counts in the peak was found using the $L_{\rm{max}}(S)$ profile, where $L_{\rm{max}}(S)$ is
the maximal value of $L$ for fixed $S$ while all others parameters were free. The distribution of $L_{\rm{max}}(S)$ values obtained from the MC
simulations for $S \geq 0$ was used to determine confidence levels in $L_{\rm max}(S)$. The limits obtained on the number of events for different
processes are shown in table 1.

\begin{table}[!htbp] \label{Table:Upper_Limits}
\begin{center}
\caption{The upper limits on the number of axions registered in Borexino FV (counts/536 days). CC - Compton axion to photon conversion,
$A+e\rightarrow e+\gamma$; AE - axio-electric effect, $A+e+Z\rightarrow e+Z$; PC - Primakoff conversion on nuclei, $A+^{12}C\rightarrow\gamma+^{12}C$.
The limits are given at 68(90)$\%$ c.l.}
\begin{tabular}{|l|c|c|c|c|}
      \hline
    reaction& CC & AE & $A$$\rightarrow$2$\gamma$ & PC     \\ \hline
     $S^{\rm lim}$    & 3.8 (6.9)  &  3.4 (6.5)   & 4.8 (8.4)   &  3.8 (6.9)       \\
\hline
\end{tabular}
\end{center}
\end{table}
The limits obtained ($S^{\rm lim}_{CC} \simeq $ 0.013 c/(100 t day) at 90\% c.l.) are very low, e.g. $\sim 10^4$ times lower than expected number of
events from $pp-$ neutrino (135 c/(100 t day)). The upper limits on the number of events with energy 5.5 MeV constrain the product of axion flux
$\Phi_A$ and the interaction cross section  with electron, proton or carbon nucleus $\sigma_{A-e,p,C}$ via
\begin{equation}
S_{\rm events} = \Phi_{A}\sigma_{A-e,p,C}N_{e,p,C}T\varepsilon \leq S^{\rm lim} \label{Events},
\end{equation}
where $N_{e,p,C}$ is the number of electrons, protons and carbon nuclei in the IV, T is the measurement time and $\varepsilon$ is
the detection efficiency. The individual rate limits are:
\begin{eqnarray}
\Phi_{A}\sigma_{A-e} \leq 4.5\times 10^{-39} \rm{s^{-1}}\\  \label{SNU_e} \Phi_{A}\sigma_{A-p} \leq 2.5\times 10^{-38}
 \rm{s^{-1}}\\\label{SNU_p}
 \Phi_{A}\sigma_{A-C} \leq 3.3\times 10^{-38} \rm{s^{-1}}.\label{SNU_C}
\end{eqnarray}
These limits show very high sensitivity to a model-independent value $\Phi_{A}\sigma_{A}$. For comparison the standard solar
neutrino capture rate is SNU = $10^{-36}\rm{s^{-1}} \rm{atom^{-1}}$. A capture rate of solar neutrinos measured by Ga-Ge
radiochemical detectors is about 70 SNU.

\subsection{Limits on $g_{Ae}$ and $g_{AN}$ couplings}

The number of expected events due to Compton conversion in the FV of the detector are:
\begin{equation}
S_{CC} = \Phi_{\nu pp}(\omega_A/\omega_{\gamma})\sigma_{CC}N_{e}T\varepsilon \label{Counts_CC}
\end{equation}
where $\sigma_{CC}$ is the Compton conversion cross sections, $\Phi_A = \Phi_{\nu pp}(\omega_A/\omega_{\gamma})$ is the axion flux (\ref{FluxA}),
$N_{e} = 9.17\times 10^{31}$ is the number of electrons in the IV; $T = 4.63\times10^7$ s is the exposure time; and $\varepsilon =0.358 $ is the
detection efficiency obtained with MC simulations (Fig.\ref{Figure:Response_Functions}).

The axion flux $\Phi_A$ is proportional to the constant $(g_{3AN})^2$, and the cross section $\sigma_{CC}$ is proportional to
the constant $g^2_{Ae}$, according to expressions (\ref{FluxA}) and (\ref{CC_CS}). The $S_{CC}$ value depends, then, on the
product of the axion-electron and axion-nucleon coupling constants: $g^2_{Ae}\times (g_{3AN})^2$. According to Eqs.
(\ref{FluxA}) and (\ref{CC_CS_Aprox}), and taking into account the approximate equality of the momenta of the axion and the
$\gamma$-quantum ($(p_A/p_{\gamma})^3 \simeq 1$ for $m_A\leq 1$ MeV), the expected number of events can be written as:
\begin{eqnarray}
\nonumber S_{CC} = g^2_{Ae}\times g^2_{3AN}\times1.4\times 10^{-14}N_{e}T\varepsilon= \\ = g^2_{Ae}\times
g^2_{3AN}\times2.1\times 10^{25}. \label{Counts_CC_Aprox}
\end{eqnarray}

\begin{figure}
\includegraphics[width=9cm,height=10.5cm]{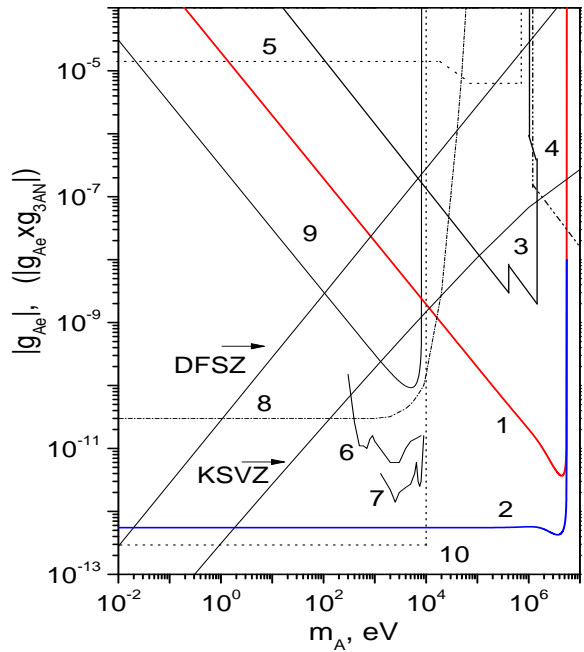}
\caption {The limits on the $g_{Ae}$ coupling constant obtained by 1- present work, 2 - present work for $|g_{Ae}\times g_{3AN}|$,  3- reactor
\cite{Alt95,Cha07} and solar experiments \cite{Bel08,Der10}, 4- beam dump experiments \cite{Kon86,Bjo88}, 5- ortho-positronium decay \cite{Asa91}, 6-
CoGeNT \cite{Aal08}, 7- CDMS \cite{Ahm09}, 8- solar axion luminosity \cite{Gon09}, 9-resonance absorption \cite{Der11},  10- read giant \cite{Raf08}.
The excluded values are located above the corresponding lines. The relations between $g_{Ae}$ and $m_A$ for KSVZ- and DFSZ-models are shown also.}
\label{Figure:gae_limits}
\end{figure}

Using this relationship, the experimental $S_{CC}^{\rm lim}$ can be used to constrain $g_{Ae}\times g_{3AN}$ and $m_A$. The range of excluded
$|g_{Ae}\times g_{3AN}|$ values is shown in Fig.\ref{Figure:gae_limits} (line 2). At $(p_A/p_{\gamma})^3 \approx 1$ or $m_A < 1 \rm{MeV}$ the limit
is:
\begin{equation}
|g_{Ae}\times g_{3AN}| \leq 5.5\times10^{-13}   ~\rm{(90\% ~c.l.).} \label{gae_gan_limit}
\end{equation}
The dependence of $|g_{Ae}\times g_{3AN}|$ on $m_A$ arises from the kinematic factor in equations (\ref{ratio}) and (\ref{CC_CS}); thus,these
constraints are completely model-independent and valid for any pseudoscalar particle. It's important to stress that the limits were obtained on the
assumption that axions escape from the Sun and reach the Earth, which implies $g_{Ae} < 10^{-6}$ for $m_A < 2m$ and $g_{Ae} < (10^{-11}-10^{-12})$ if
$m_A > 2m$ (\cite{Der10}).

Within the hadronic (KSVZ) axion model, $g_{3AN}$  and $m_A$ are related by expression (\ref{gan3}), which can be used to obtain
a constraint on the $g_{Ae}$ constant, depending on the axion mass (Fig.\ref{Figure:gae_limits}. line 1). For
$(p_A/p_{\gamma})^3 \approx 1$ the limit on $g_{Ae}$ and $m_A$ is:
\begin{equation}
|g_{Ae}\times m_A| \leq 2.0\times10^{-5} \rm{~eV}  ~\rm{(90\% ~c.l.),} \label{gae_ma_limit}
\end{equation}
where $m_A$ is given in eV units. For $m_A$ = 1 MeV, this constraint corresponds to $g_{Ae}\leq 2.0 \times 10^{-11}$. Figure \ref{Figure:gae_limits}
shows the constraints on $g_{Ae}$ that were obtained in experiments with reactor, accelerator, and solar axions, as well as constraints from
astrophysical arguments.

\subsection{Limits on $g_{A\gamma}$ and $g_{AN}$ couplings}
The analysis of $A\rightarrow 2\gamma$ decay and Primakoff photoproduction is more complicated because axions can decay during their flight from the
Sun. The exponential dependence of the axion flux on $g_{a\gamma}$ and $m_A$, given by (\ref{Flux}), must be taken into account.

The number of events detected in the FV due to axion decays into 2
$\gamma$'s within the IV are:
\begin{equation}
S_{2\gamma} = N_{\gamma}T \varepsilon_{2\gamma} \label{Counts_2G} 
\end{equation}
where $N_{\gamma}$ is given by (\ref{N_gamma}) and $\varepsilon_{2\gamma}$ = 0.35 is the detection efficiency obtained by MC simulation. The relation
$S_{2\gamma} < S^{\rm lim}_{A\rightarrow 2\gamma}$ leads to model-independent limits on $g_{3AN}^2\times g_{A\gamma}^2$ vs axion mass. The expected
value of $S_{2\gamma}$ has a complex dependence on $g_{A\gamma}$, $g_{3AN}$ and $m_A$ given by equations (\ref{tau_CM})-(\ref{N_gamma}).

In the assumption that $\beta \approx 1$ the number of decays in the FV depends on $g_{3AN}^2$, $g_{A\gamma}^2$ and $m_A^4$:
\begin{equation}
N_{\gamma}= 1.68\times 10^{-4}g^2_{A\gamma} \times g_{3AN}^2\times m_A^4, \label{NG}
\end{equation}
where $g_{A\gamma}$ and $m_A$ are given in $\rm{GeV}^{-1}$  and eV units, respectively. The limit derived from equation
(\ref{Counts_2G}), at 90\% c.l., is
\begin{equation}
 |g_{A\gamma}\times g_{3AN}|\times m_A^2 \leq 3.3\times 10^{-11} \rm{~eV}. \label{AGMA}
\end{equation}

\begin{figure}
\includegraphics[width=9cm,height=10.5cm]{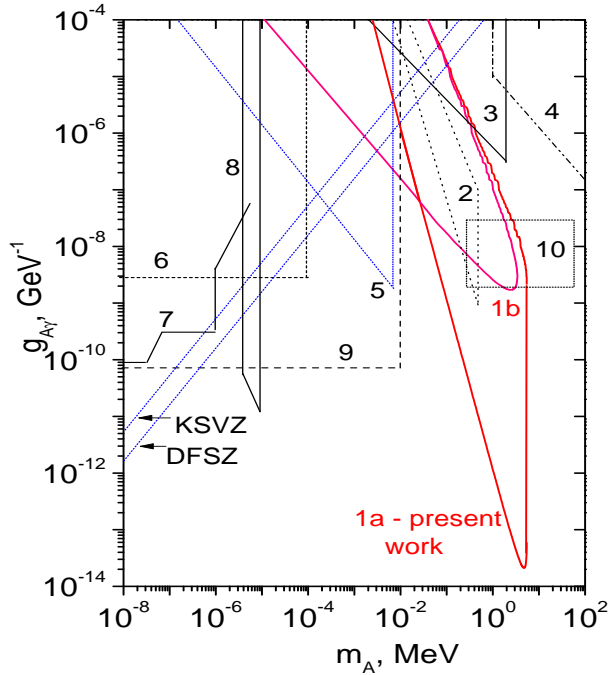}
\caption {The limits on $g_{A\gamma}$ obtained by 1- present work (a - $A\rightarrow 2\gamma$, b - PC, areas of excluded values are located inside
contour), 2 - CTF \cite{Bel08}, 3- reactor experiment \cite{Cha07}, 4- beam dump experiments \cite{Kon86,Bjo88}, 5- resonant absorption \cite{Der09},
6- solar axions conversion in crystals - \cite{Avi99,Ber01,Mor02}, 7- CAST and Tokyo helioscope \cite{Zio05,Ari09,Ino08}, 8-telescopes
\cite{Ber91,Res91,Gri07}, 9- HB Stars \cite{Raf08}, 10- expectation region from heavy axion models \cite{Bere00,Bere01,Hal04}. }
\label{Figure:gag_limits}
\end{figure}

The dependence of $S_{2\gamma}$ on $g_{A\gamma}$ and $m_A$ is obtained from (\ref{gan3}),  which gives the relationship between $g_{3AN}$ and $m_A$
in the KSVZ model. The relation $S_{2\gamma} \leq S^{\rm lim}_{A\rightarrow 2\gamma}$ imposes constraints on the range of $g_{A\gamma}$ and $m_A$
values. The excluded region is inside contour 1a in Fig.\ref{Figure:gag_limits} (90 \% c.l.). For higher values of $g_{A\gamma}^2m_A^3$ axions decay
before they reach the detector, while for lower $g_{A\gamma}^2m_A^3$ the probability of axion decay inside the Borexino volume is too low.  The
limits on $g_{A\gamma}$ obtained by other experiments are also shown.

The Borexino results exclude a large new region of axion-photon coupling constant $(2\times10^{-14} - 10^{-7}) \rm{GeV}^{-1}$ for the axion mass
range $(0.01 - 5)$ MeV. The Borexino limits are about 2-4 order of magnitude stronger than those obtained by laboratory-based experiments using
nuclear reactors and accelerators. Moreover, our excluded region has begun to overlap the predicted regions from heavy axion models
\cite{Bere00,Bere01, Hal04}.

At $m_A < 1 \rm{MeV}$ the constraint on $g_{A\gamma}$ and $m_A$ is given by
\begin{equation}
 |g_{A\gamma}|\times m_A^3 \leq 1.2\times 10^{-3} \rm{~eV^2}. \label{AGMA}
\end{equation}
So, e.g., $m_A$=1 MeV corresponds to $g_{A\gamma}\leq1.2\times 10^{-12}\rm{~GeV^{-1}}$. Under the assumption that the axion-photon coupling
$g_{A\gamma}$ depends on axion mass as in the KSVZ model (\ref{C_gamma}), we exclude axions with mass in the (7.5 - 76) keV range (see
Fig.\ref{Figure:Decay_PC}). Similar constraints can be obtained for DFSZ axions for specific values of $\cos^2\beta_{\rm{dfsz}}$.

The number of expected events due to inverse Primakoff conversion is:
\begin{equation}
S_{PC} = \Phi_A\sigma_{PC}N_{C}T\varepsilon_{PC} \label{Counts_PC}
\end{equation}
where $\sigma_{PC}$ is the Primakoff conversion cross sections; $N_{C}$ is the number of  carbon nuclei in the IV, and $\varepsilon_{PC}$ is the
detection efficiency for 5.5 MeV $\gamma$'s. The axion flux, $\Phi_A$, is proportional to the constant $g^2_{3AN}$, and the cross section
$\sigma_{PC}$ is proportional to the constant $g^2_{A\gamma}$, according to equations (\ref{FluxA}) and (\ref{Primakoff}). As a result, the $S_{PC}$
value depends on the product of the axion-photon and axion-nucleon coupling constants: $g^2_{A\gamma}\times g^2_{3AN}$. Under the assumption  that
$\Phi_A \approx \Phi_{A0}$ (true for $g_{A\gamma}(\rm{GeV^{-1}}) \times m_A^2 \rm{(eV)}< 1.2\times10^4$) one can obtain the limit:

\begin{equation}
 |g_{A\gamma}\times g_{3AN}|\leq 4.6\times 10^{-11}\rm{~GeV^{-1}} ~{\rm (90\%~c.l.),} \label{AAGPC}
\end{equation}
where again $g_{A\gamma}$ is in GeV$^{-1}$ units. This limit is 25 times stronger than the one obtained  by CAST \cite{And10},
which searches for conversion of 5.5 MeV axions in a laboratory magnetic field ($|g_{A\gamma}\times g_{3AN}|\leq 1.1\times
10^{-9}$ at $m_A \leq 1 \rm{eV}$).

In the KSVZ model (\ref{gan3}), the constraint on  $g_{A\gamma}$ and $m_A$ is given by the relation:
\begin{equation}
 |g_{A\gamma}|\times m_A\leq 1.7\times 10^{-12}. \label{AGMAPC}
\end{equation}
For $m_A$=1 MeV, this corresponds to $g_{A\gamma}\leq1.7\times 10^{-9}\rm{~GeV^{-1}}$. The region of excluded values of
$g_{A\gamma}$ and $m_A$ are shown in Fig.\ref{Figure:gag_limits}, line 1b; under the assumption that $g_{A\gamma}$ depends on
$m_A$ as in the KSVZ model (\ref{C_gamma}) we exclude axions with masses between (1.5 - 73) keV (see Fig.\ref{Figure:Decay_PC}).
Our results from the inverse Primakoff process exclude a new region of $g_{A\gamma}$ values at $m_A \sim$ 10 keV.

\section{CONCLUSIONS}
A search for 5.5 MeV solar axions emitted in the $p(d,\rm{^3He})A$ reaction has been performed with the Borexino detector. The
Compton conversion of axions into photons, the decay of axions into two photons, and inverse Primakoff conversion on nuclei were
studied. The signature of all these reactions is a 5.5 MeV peak in the energy spectrum of Borexino. No statistically significant
indications of axion interactions were found. New, model independent, upper limits on the axion coupling constants to electrons,
photons and nucleons,
\begin{equation}
 |g_{Ae}\times g_{3AN}|\leq 5.5\times10^{-13}
\end{equation}
and
\begin{equation}
 |g_{A\gamma}\times g_{3AN}|\leq4.6\times 10^{-11} \rm{GeV^{-1}},
\end{equation}
were obtained at $m_A < 1 \rm{MeV}$ and 90\% c.l.

Under the assumption that $g_{3AN}$ depends on $m_A$ as in the KSVZ axion model, new 90\% c.l. limits on axion-electron and
axion-photon coupling as a function of axion mass were obtained:
\begin{equation}
|g_{Ae}\times m_A| \leq 2.0\times10^{-5}\rm{~eV}
\end{equation}
and
\begin{equation}
 |g_{A\gamma}\times m_A|\leq 1.7\times 10^{-12}.
\end{equation}
The new Borexino results exclude large regions of axion-electron and axion-photon coupling constants ($g_{Ae} \in (10^{-11} - 10^{-9})$ and
$g_{A\gamma} \in (2\times10^{-14} - 10^{-7}) \rm{GeV}^{-1}$) for the axion mass range $(0.01 - 5)$ MeV.

\section{ACKNOWLEDGMENTS}
The Borexino program was made possible by funding from INFN and PRIN 2007 MIUR (Italy), NSF (USA), BMBF, DFG, and MPG (Germany), NRC Kurchatov
Institute (Russia), and MNiSW (Poland). We acknowledge the generous support of the Laboratori Nazionali del Gran Sasso (LNGS). A.~Derbin, L.~Ludhova
and O.~Smirnov acknowledge the support of Fondazione Cariplo.

\end{document}